\documentclass[a4paper, 10pt, conference]{IEEEtran}
\IEEEoverridecommandlockouts
\usepackage{cite}
\usepackage{amsmath,amssymb,amsfonts}
\usepackage{algorithmic}
\usepackage{graphicx}
\usepackage{textcomp}
\usepackage{xcolor}
\usepackage{comment}
\usepackage{csquotes}
\usepackage{hyperref}
\usepackage{tikz}
\usetikzlibrary{positioning}

\usepackage{ragged2e}
\usepackage{subcaption}

\usepackage{multirow}

\makeatletter % changes the catcode of @ to 11
\newcommand{\linebreakand}{%
  \end{@IEEEauthorhalign}
  \hfill\mbox{}\par
  \mbox{}\hfill\begin{@IEEEauthorhalign}
}
\makeatother % changes the catcode of @ back to 12

\def\BibTeX{{\rm B\kern-.05em{\sc i\kern-.025em b}\kern-.08em
    T\kern-.1667em\lower.7ex\hbox{E}\kern-.125emX}}
\begin{document}

%\title{Exploring the Effect of Heights on \\ User Experience in Extended Reality}
\title{Exploring the Effect of Heights and User Stance on User Experience in Extended Reality Climbing}

%Emotion Elicitation in Remote VR Studies: Continuous Ratings vs. Post-Questionnaires


\author{
 \IEEEauthorblockN{Tanja Koji\'c$^1$, Nathan Kirchner$^1$, Maurizio Vergari$^1$, Maximilian Warsinke$^1$,  \\ Sebastian M\"oller$^{1,2}$, Jan-Niklas Voigt-Antons$^3$}
 \IEEEauthorblockA{$^1$Quality and Usability Lab, Technische Universität Berlin, Germany, \\ $^2$German Research Center for Artificial Intelligence (DFKI), Berlin, Germany\\ $^3$Immersive Reality Lab, Hamm-Lippstadt University of Applied Sciences, Hamm, Germany}
}

\maketitle

\begin{abstract}
Virtual environments (VEs) are increasingly used for immersive experiences, training simulations, and entertainment, yet factors such as height perception and user stance can significantly influence user experience (UX). Height perception in VEs plays a crucial role in shaping UX, particularly in immersive applications such as climbing simulations. This study investigates the effects of height in various VEs and examines how user stance, sitting or standing, impacts immersion, perceived height, and motion sickness. A user study was conducted with 25 participants who played through five randomized climbing scenarios, ranging from indoor climbing gyms to outdoor cityscapes and mountainous terrains. Participants’ UX was assessed using standardized questionnaires, including the IPQ for general presence, spatial presence, involvement, and experienced realism; as well as the SSQ to evaluate motion sickness symptoms such as nausea, oculomotor strain, and disorientation. Results indicate that seated participants experienced slightly higher immersion but were also more susceptible to motion sickness compared to those standing. While standing participants maintained consistent scores across different environments, seated participants reported increased immersion and discomfort as the VEs became larger, more physically demanding, and visually complex. %These findings contribute to a broader understanding of how physical positioning and environmental complexity shape user experience in XR applications, with implications for the design of more comfortable and engaging virtual experiences.
\end{abstract}

\begin{comment}
\newcommand\copyrighttext{%
  \footnotesize \textcopyright\ 2024 IEEE. Personal use of this material is permitted. Permission from IEEE must be obtained for all other uses, in any current or future media, including reprinting/republishing this material for advertising or promotional purposes, creating new collective works, for resale or redistribution to servers or lists, or reuse of any copyrighted component of this work in other works. The original version of this article is available at: \href{https://ieeexplore.ieee.org/document/10598301}{https://ieeexplore.ieee.org/document/10598301} DOI:\href{https://doi.org/10.1109/QoMEX61742.2024.10598301}{10.1109/QoMEX61742.2024.10598301}%
}
\end{comment}

\newcommand\copyrighttext{%
    \footnotesize \textcopyright 2026 IEEE. Personal use
    of this material is permitted. Permission from IEEE
    must be obtained for all other uses, in any current or
    future media, including reprinting/republishing this
    material for advertising or promotional purposes,
    creating new collective works, for resale or
    redistribution to servers or lists, or reuse of any
    copyrighted component of this work in other works.
    https://doi.org/10.1109/QoMEX65720.2025.11219977}

\newcommand\copyrightnotice{%
\begin{tikzpicture}[remember picture,overlay,shift=
    {(current page.south)}]
    \node[anchor=south,yshift=10pt] at (0,0)
    {\fbox{\parbox{\dimexpr\textwidth-\fboxsep-
    \fboxrule\relax}{\copyrighttext}}};
\end{tikzpicture}%
}
\copyrightnotice


%\begin{tikzpicture}[overlay, remember picture]
%\path (current page.north) node (anchor) {};
%\node [below=of anchor] {%
%2024 16th International Conference on Quality of Multimedia Experience (QoMEX)};
%\end{tikzpicture}

\section{Introduction}

Height perception plays a critical role in shaping user experience (UX) in Extended Reality (XR), particularly in applications involving vertical navigation, such as climbing, construction, or exploration \cite{asjad2018perception}. While XR enables the simulation of height-related tasks, it can also elicit discomfort such as simulator sickness due to sensory mismatch, postural strain, or environmental complexity. Prior work has emphasized factors such as visual fidelity, spatial layout, and motion cues in shaping presence and immersion \cite{chang2020virtual, pastel2022comparison}, but comparatively less attention has been given to how users' physical stance, sitting or standing, modulates these effects during vertical tasks.
Understanding the relationship between user stance and virtual height exposure is important for designing XR applications that balance realism, comfort, and immersion \cite{kim2020multisensory}. For instance, seated users may feel physically safer but perceptually more disconnected from vertical movement. Standing users, on the other hand, may experience more embodied engagement but face greater postural demand.

This paper investigates how user stance affects the perceived realism, immersion, and simulator sickness in VR climbing tasks. %To this end, a user study was conducted with 25 participants across five virtual climbing environments ranging from indoor gyms to tall cityscapes and mountainous landscapes. Participants experienced the scenarios while alternating between sitting and standing conditions.

\newpage
The study addresses the following research questions:
\begin{itemize}
    \item How do variations in virtual environment design influence users' perception of height in XR?
    \item How does user stance (sitting vs. standing) affect overall user experience and comfort in XR?
\end{itemize}

\vspace{0.5em}
%Our findings reveal trends suggesting that seated users report slightly higher realism but also experience increased simulator sickness in complex environments, while standing users maintain more stable levels of comfort. These insights contribute to XR design by identifying trade-offs between embodiment and discomfort under different environmental and postural conditions.
In this study, we aim to find out whether sitting makes users feel that the virtual environment is more realistic, but also causes more simulator sickness, especially in complex scenes. We also look at whether standing helps users feel more comfortable overall. By exploring these differences, we want to better understand how body posture and environment design affect the XR experience.

\section{Related Work}

Research on UX in XR has shown that factors such as spatial fidelity, locomotion method, and postural engagement can significantly influence presence and comfort. Prior work on height perception in XR suggests that visual elevation alone does not linearly correlate with fear responses or physiological arousal \cite{fear-of-heights}. However, immersive climbing tasks can amplify users’ awareness of height, making them valuable for examining perceptual and physiological responses to verticality.

Cybersickness is another well-documented issue in XR, often attributed to sensory mismatches between visual and vestibular cues \cite{chang2020virtual}. Studies have shown that simulator sickness symptoms such as nausea and disorientation increase in visually complex or motion-intensive environments\cite{motion-sickness}, especially when interaction techniques involve artificial locomotion \cite{kumar2024effects}. However, less is known about how static stances, like sitting vs. standing, impact these symptoms during physically grounded interactions such as climbing.

Postural configuration itself has been shown to affect presence, with standing often yielding greater embodiment and environmental awareness, while seated use may reduce physical strain but increase perceptual disconnect \cite{zielasko2021sit}. These trade-offs are especially relevant in VR climbing scenarios, where user movement is localized, but the perception of height is simulated.

While exposure therapy studies have explored fear reduction through controlled virtual height environments \cite{exposure-therapy}, their objectives differ from this work. We do not assess therapeutic outcomes, but rather examine how posture and environmental design influence non-clinical user experience in height-based XR tasks.
Our work builds on this literature by systematically varying user stance across multiple vertical environments and assessing its effects on presence and discomfort using standardized UX measures.

\section{Methods}
\subsection{Study Design}

\begin{figure}[h]
    \centering
    \begin{subfigure}[b]{0.317\textwidth}
        \centering
        \includegraphics[height=3.1cm,keepaspectratio]{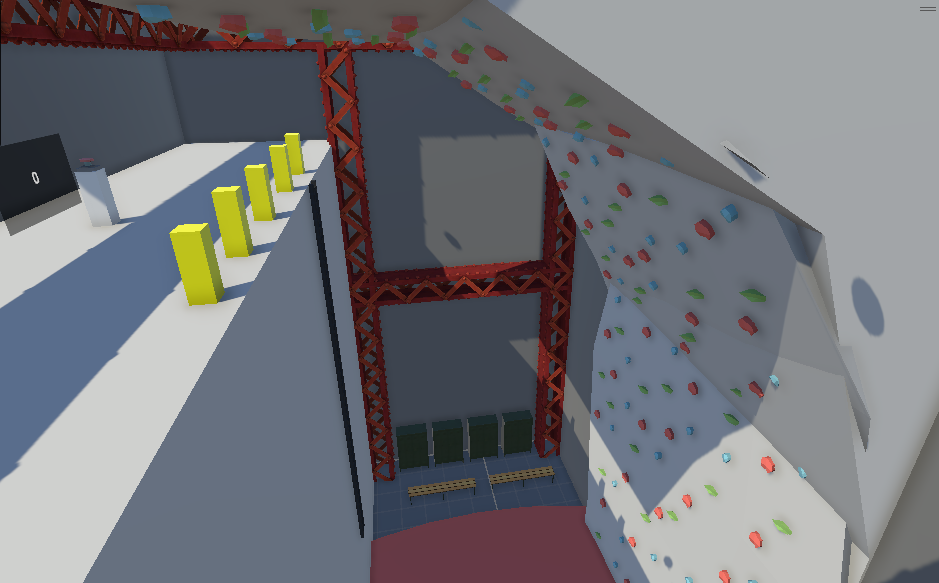}
        \subcaption{Climbing Gym}
        \label{lvl:Climbing Gym}
    \end{subfigure}
    \begin{subfigure}[b]{0.317\textwidth}
        \centering
        \includegraphics[height=3.1cm,keepaspectratio]{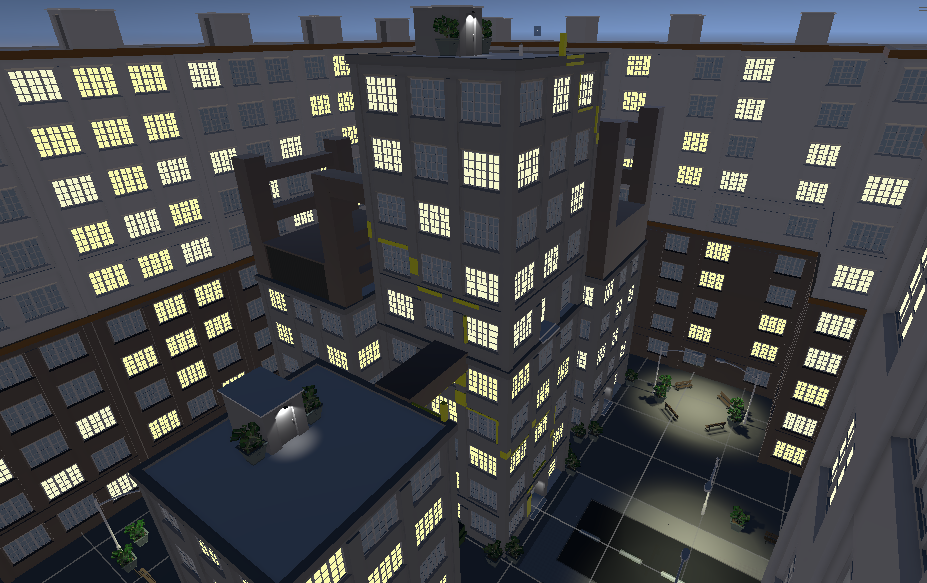}
        \subcaption{City Nighttime}
        \label{lvl:City Nighttime}
    \end{subfigure}
    \begin{subfigure}[b]{0.317\textwidth}
        \centering
        \includegraphics[height=3.1cm,keepaspectratio]{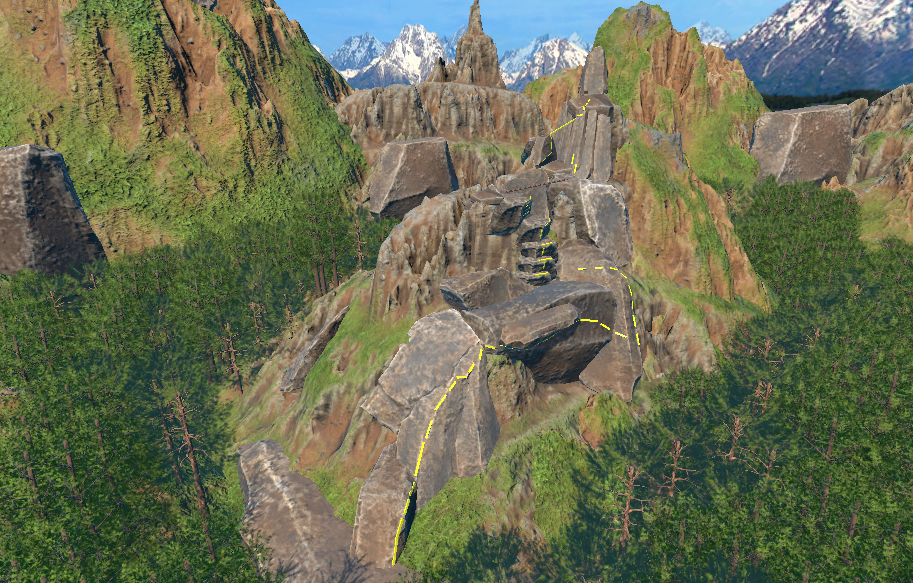}
        \subcaption{Mountain}
        \label{lvl:Mountain}
    \end{subfigure}  
    \caption{Different levels in Unity climbing application}
\end{figure}

We developed a climbing simulation in Unity3D using the Meta Quest 3 headset to investigate how user stance affects experience in height-based virtual environments. Participants completed climbing tasks in five distinct virtual environments (VEs), each designed to represent a different level of visual fidelity and height exposure. The environments were: \textit{Climbing Gym 1} and \textit{Climbing Gym 2} (differing in wall complexity and hold density), \textit{City Rooftop Day}, \textit{City Rooftop Night}, and \textit{Mountain Summit}. These five levels offered structured variation in elevation, lighting, and spatial layout complexity (see \ref{lvl:City Nighttime}, \ref{lvl:Climbing Gym}, \ref{lvl:Mountain}).
All interaction was performed through direct physical climbing by grabbing virtual holds using XR controllers. No artificial locomotion methods such as joystick navigation or teleportation were used. This approach aimed to isolate the effects of environment and user stance without introducing visual-vestibular conflict from indirect movement.
Each participant experienced all five levels in a randomized order. The stance condition (sitting or standing) alternated between levels, ensuring that each participant completed two or three levels per condition. While this reduced order effects, it introduced slight imbalance across conditions. A stopwatch was used to track level completion time to support consistent pacing.

\subsection{Participants}

A total of 25 participants (13 female, 12 male; age M = 26.4, SD = 7.6) took part in the study. Twelve participants reported no prior experience with XR systems, and none identified as experienced climbers. Participants were recruited through mailing lists and word of mouth. 
No exclusion criteria were applied regarding physical fitness, gaming experience, or height sensitivity. However, participants were informed about the nature of the climbing simulation and potential risk of discomfort prior to the study. They were monitored closely for signs of simulator sickness and given the option to withdraw at any point. All participants completed the full procedure without interruption. Written informed consent was obtained from all participants in accordance with institutional ethics guidelines.

\subsection{Procedure and Measures}

After a tutorial level, participants completed the five climbing environments with alternating stance conditions. Each level lasted approximately 3–5 minutes. After each level, participants removed the headset and completed two standardized questionnaires:

\begin{itemize}
    \item \textbf{IPQ} \cite{IPQ}: to assess spatial presence, involvement, and realism.
    \item \textbf{SSQ} \cite{SSQ}: to measure symptoms of simulator sickness (nausea, oculomotor strain, disorientation).
\end{itemize}

Participants could pause or withdraw if they experienced discomfort. A final open-ended questionnaire collected qualitative impressions about realism, comfort, and perceived height.

\subsection{Analysis}

A repeated-measures ANOVA was conducted to examine the effects of environment and stance on IPQ and SSQ subscales. Where applicable, Mauchly’s test for sphericity and Shapiro–Wilk tests for normality were conducted; if assumptions were violated, non-parametric alternatives (Friedman tests) were applied. Statistical significance was set at $\alpha = .05$.

\section{Results}

\subsection{Realism and Nausea}

\begin{figure}[ht!]
    \centering
    \includegraphics[width=0.9\linewidth]{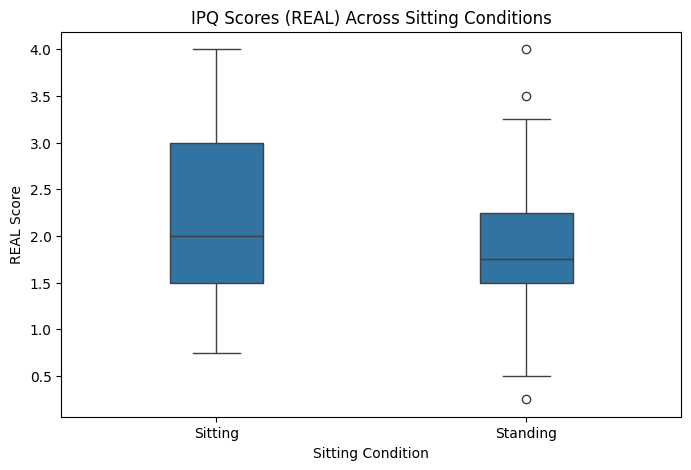}
    \caption{IPQ realism (REAL) scores across sitting conditions}
    \label{fig:real}
    \vspace{-1em}
\end{figure}

A repeated-measures ANOVA revealed a significant effect of user stance on the IPQ realism subscale (\textit{F}(1, 114) = 4.79, $p$ = .031), with seated participants reporting higher realism (M = 2.24, SD = 0.95) than those standing (M = 1.87, SD = 0.84), as shown in Fig.~\ref{fig:real}. No significant differences were found for other IPQ subscales.

\begin{figure}[ht!]
    \centering
    \vspace{-1em}
    \includegraphics[width=0.9\linewidth]{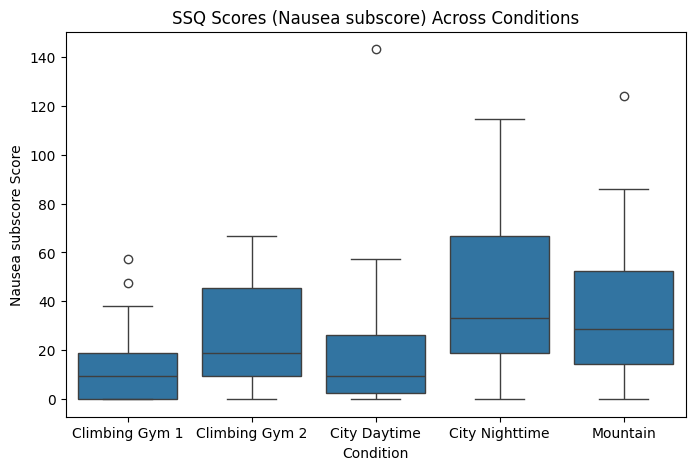}
    \caption{SSQ nausea subscore across conditions}
    \label{fig:nausea}
        \vspace{-1em}
\end{figure}

Simulator sickness symptoms varied significantly by environment, particularly for nausea (\textit{F}(4, 111) = 4.78, $p$ = .001). As shown in Fig.~\ref{fig:nausea}, nausea levels peaked in the City Nighttime condition (M = 43.3, SD = 31.1), followed by the Mountain scene. The lowest nausea scores were observed in Climbing Gym 1 (M = 13.4, SD = 16.3). Post hoc tests confirmed significant differences between City Nighttime and Climbing Gym 1 ($p$ = .0014), as well as between Mountain and Climbing Gym 1 ($p$ = .0417).

\subsection{Overall SSQ Trends}

\begin{figure}[ht!]
    \centering
    \includegraphics[width=0.7\linewidth]{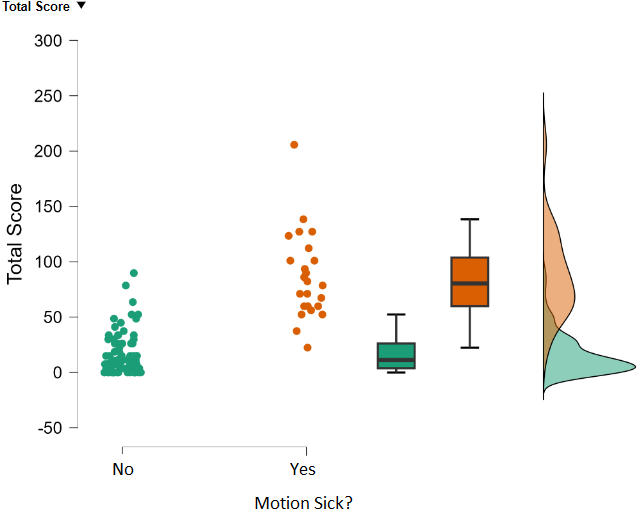}
    \caption{Total SSQ Score split by felt motion sickness}
    \label{fig:total_score_motion_sick}
    \vspace{-1em}

\end{figure}

Total SSQ scores, shown in Fig.~\ref{fig:total_score_motion_sick}, did not differ significantly by stance, but seated participants displayed greater score variability and reported higher discomfort in visually complex environments. Standing participants maintained relatively stable sickness levels across all scenes.

\subsection{Learning and Individual Differences}

Due to the randomized scene order, some levels were played more frequently in certain stance conditions at specific times, limiting direct comparison. Variability in individual SSQ scores (some users experienced symptoms at low scores, others reported none at high scores) further emphasizes the subjective nature of XR discomfort.

\section{Discussion}

The results indicate that user stance subtly influences user experience in XR climbing and similar vertical scenarios. Seated participants reported higher realism but also more simulator sickness, especially in visually rich scenes like City Nighttime and Mountain, suggesting a comfort–immersion trade-off. Standing participants, by contrast, showed more stable comfort levels, which may benefit motion-sensitive users.

These findings suggest that physical posture may affect how users process environmental stimuli and vertical exposure. Prior work has shown that postural engagement enhances spatial awareness and presence \cite{zielasko2021sit}, while seated configurations may reduce postural strain but increase susceptibility to sensory mismatch \cite{kumar2024effects}. Our results align with these insights, showing greater sickness variance in seated users during complex scenes, where motion and height are emphasized.
Environmental complexity itself appears to contribute to nausea, supporting previous research that links sensory load to cybersickness \cite{chang2020virtual}. However, since participants interacted through physical climbing rather than artificial locomotion, we attribute these symptoms primarily to visual and spatial complexity rather than movement technique. That said, we cannot fully rule out the contribution of prolonged exposure or subtle mismatches between virtual height and body cues.
Subjective feedback reinforced the quantitative results, with participants perceiving the taller, open environments as more immersive but also more demanding. Observations, such as hesitation during wooden plank crossings, suggest that simulated height can provoke instinctive reactions, even in non-clinical samples.
Importantly, the randomized design caused slight imbalance in stance exposure, as some levels appeared more often in seated or standing order. This limits strict comparability and should be addressed in future counterbalanced designs. This limits our ability to draw strong statistical conclusions and highlights the need for counterbalanced designs in future.

\section{Conclusion}

This study explored how user stance and virtual environment design influence user experience in XR climbing scenarios. Results suggest that seated users reported slightly higher realism but were more susceptible to simulator sickness, particularly in visually complex scenes. Standing participants showed more consistent comfort levels across environments. While these trends indicate a trade-off between immersion and discomfort, the study is limitations of a small sample and not full randomization constrain generalizability and should be addressed in future research. %Future work should incorporate counterbalanced designs and physiological measures to better understand how postural configuration and environmental complexity interact to shape XR comfort and immersion.

\bibliographystyle{IEEEtran}
\bibliography{IEEEabrv, main}
%\nocite{*}
\end{document}